\documentclass[12pt]{iopart}
\pdfoutput=1
\usepackage{iopams}
 
\expandafter\let\csname equation*\endcsname\relax
\expandafter\let\csname endequation*\endcsname\relax
\usepackage{amsmath,amssymb,mathtools}

\usepackage{graphicx,color}
\usepackage{enumerate}

\usepackage[colorlinks,linkcolor=blue,urlcolor=blue,citecolor=blue]{hyperref}

 \newcommand{\erefss}[1]{\eqref{#1}}%
 \newcommand{\erefs}[1]{\eqref{#1}}%
 \newcommand{\frefss}[1]{figures~\ref{#1}} %
 \newcommand{\frefs}[1]{~\ref{#1}} %
 \newcommand{\ssref}[1]{Section~\ref{#1}}%
 \newcommand{\aref}[1]{\ref{#1}}%

\newcommand\be{\begin{equation}} 
\newcommand\ee{\end{equation}} 
\begin{document}

\title{Thermodynamic uncertainty relations  in a linear system}

\author{Deepak Gupta and Amos Maritan}  
\address{Dipartimento di Fisica `G. Galilei', INFN, Universit\`a di Padova, Via Marzolo 8, 35131 Padova, Italy}

\date{\today}

\begin{abstract}
We consider a Brownian particle in harmonic confinement of stiffness $k$, in one dimension in the underdamped regime. The whole setup is immersed in a heat bath at temperature $T$. The center of harmonic trap is dragged under any arbitrary protocol. The thermodynamic uncertainty relations for both position of the particle and current at time $t$ are obtained using the second law of thermodynamics as well as the positive semi-definite property of the correlation matrix of work and degrees of freedom of the system for both underdamped and overdamped cases. 
\end{abstract}


\maketitle

\markboth{Thermodynamic uncertainty relations  in a linear system}{}

\section{Introduction}
\label{intro}
Stochastic thermodynamics \cite{Seifert-2,Seifert-1,Sekimoto} provides a platform to understand properties of small-systems. These systems include Brownian particles (colloidal particles), bio-molecular motors, small-scale engines, DNA and RNA molecules, proteins, etc..   Since the length scale of such systems is small, fluctuations present in the surrounding environment perturb the deterministic nature of these systems. Therefore, they evolve under stochastic dynamics, and the evolution of the probability of a system to be in a given configuration is described by master and Fokker-Planck equations. Moreover, the observables such as work on the system, the heat exchanged by the system with the environment, entropy production, etc., can be defined along a stochastic trajectory  within the framework of stochastic thermodynamics. In the past two decades, some universal results in the nonequilibrium physics gained much attention; namely, fluctuation theorems \cite{Evans-Cohen, Evans-Searles,  Gallavotti-Cohen, Gallavotti1995, Kurchan, Lebowitz-Spohn,Searles-Evans,Searles2001}, Jarzynski equality \cite{Jarzynski-PRL}, Crooks work fluctuation theorem \cite{Crooks1998,Crooks-1,Crooks-2}, etc..

Recently, there has been an explosion of research in understanding \emph{thermodynamic uncertainty relations} which quantify the trade-off between the precision of current (particle current, heat current, electron flux in a quantum dot, etc.) and the thermodynamic cost in various systems. Suppose $\phi$ and $\sigma$ be the current and the average entropy production in a nonequilibrium process; the thermodynamic uncertainty relation relates these two observables as following:
\begin{align}
\dfrac{\text{Var}[\phi]}{(\text{Mean}[\phi])^2}\geq \dfrac{2}{\sigma},
\end{align}
where $\text{Var}[\phi]$ and $\text{Mean}[\phi]$ represent the variance and the mean of an observable $\phi$, respectively. The above relation states that to reduce the fluctuations of $\phi$ (gain more precision), one has to pay a large thermodynamic cost quantified by the average entropy production $\sigma$. It is indeed a remarkable result in the nonequilibrium statistical physics.

First thermodynamic uncertainty relation was obtained by Barato and Seifert \cite{seifert-ur-1} for bio-molecular processes for both unicyclic and multicyclic networks. Later, an extension of result \cite{seifert-ur-1} is shown for periodically driven systems \cite{Barato_2018,Koyuk_2018}, chemical kinetics \cite{seifert-ur-2,PRX}, finite time processes \cite{todd-2,Maes,Manikandan,Felix, Pigolotti-1}, counting observables \cite{Garrahan}, and biochemical sensing \cite{Seifert-ur-3}.  Gingrich \emph{et al.} \cite{todd-1} obtained a bound for the large deviation function \cite{Touchette} for steady state empirical currents in Markov jump processes and proved the thermodynamic uncertainty relation conjectured in \cite{seifert-ur-1}, and then, its tighter version is obtained in \cite{Polettini}. Several other generalization such as parabolic bound, exponential bound, hyperbolic cosine bound, etc., for currents in the nonequilibrium steady state are obtained in \cite{seifert-ur-4}. Interestingly, Shiraishi \cite{Shiraishi} showed that the original thermodynamic uncertainty  relation  \cite{seifert-ur-1,todd-1} does not hold for the discrete time Markov chain. Later, Proesmans \emph{et al.} \cite{Proesmans_2017} obtained the thermodynamic uncertainty relation for  the discrete time Markov chain using the large deviation techniques. A connection between discrete and continuous time uncertainty relations is shown in \cite{Mapping}. One can also see similar uncertainty relations in the context of discrete processes \cite{Baiesi}, multidimensional systems \cite{Dechant-2018-MD}, Brownian motion in the tilted periodic potential \cite{Hwang}, general Langevin systems \cite{Dechant_2018}, molecular motors \cite{Hwang-2}, run and tumble processes \cite{Mayank}, biochemical oscillations \cite{Marsland}, interacting oscillators \cite{Lee}, effect of magnetic field \cite{chun-magnetic}, linear response \cite{Katarzyna}, measurement and feedback control \cite{hasegawa-inf-fdbk}, information \cite{information}, underdamped Langevin dynamics \cite{underdamped}, time-delayed Langevin systems \cite{time-delay}, various systems \cite{Chetrite}, etc.. Recently, Hasegawa \emph{et al.} \cite{general} found an uncertainty relation for the time-asymmetric observable for the system driven by a time-symmetric driving protocol using the steady state fluctuation theorem.  However, to our knowledge, the similar relation for a generic current in the presence of an arbitrary external protocol in a general setup is still an open question. We aim to partially answer this question in a simple setup (described below) driven out of equilibrium using an arbitrary time-dependent external protocol. A generalization of \cite{general} for the broken time-reversal symmetry using the time-reversed current observable is given in \cite{patrick,Kerel-2019}.

In this paper, we consider a one-dimensional system of a Brownian particle confined in a harmonic trap. The whole arrangement is coupled with a heat bath at temperature $T$. The center of the harmonic trap is dragged with an arbitrary protocol $\lambda(t)$. Here, we focus on two observables: (1) the position of the particle (even variable with respect to time), (2) the current (odd variable  with respect to time) \cite{general} which measures the distance of the particle at time $t$ from the initial location at $t=0$. The uncertainty relations for both of these observables are obtained using the \emph{second law of thermodynamics} as well as the positive semi-definite property of the correlation matrix of work and degrees of freedom of the system in both underdamped and overdamped regimes. There are four main features of the paper: (1) the thermodynamic uncertainty relations are obtained from the second law of thermodynamics, (2) the external protocol need not be time-symmetric, (3) in contrary to \cite{patrick,Kerel-2019}, there is no need to compute the observable in the time-reversed protocol, (4) the cost function is given by work done on the system in the overdamped regime.

The remaining of the paper is organized as follows. In \sref{sec-model}, we discuss our model. \ssref{sec-tur} contains the derivation of the thermodynamic uncertainty relations in both underdamped and overdamped regimes. Finally, we summarized our paper in \sref{summ}. In \aref{jpdf}, we give the joint probability density function $\mathcal{P}(x,v,\mathcal{W},t)$. We show small- and large-time behaviors of the mean position of particle and the mean work on the Brownian particle in \aref{asymp}. 
\begin{figure}
  \begin{center}
    \includegraphics[width=8cm]{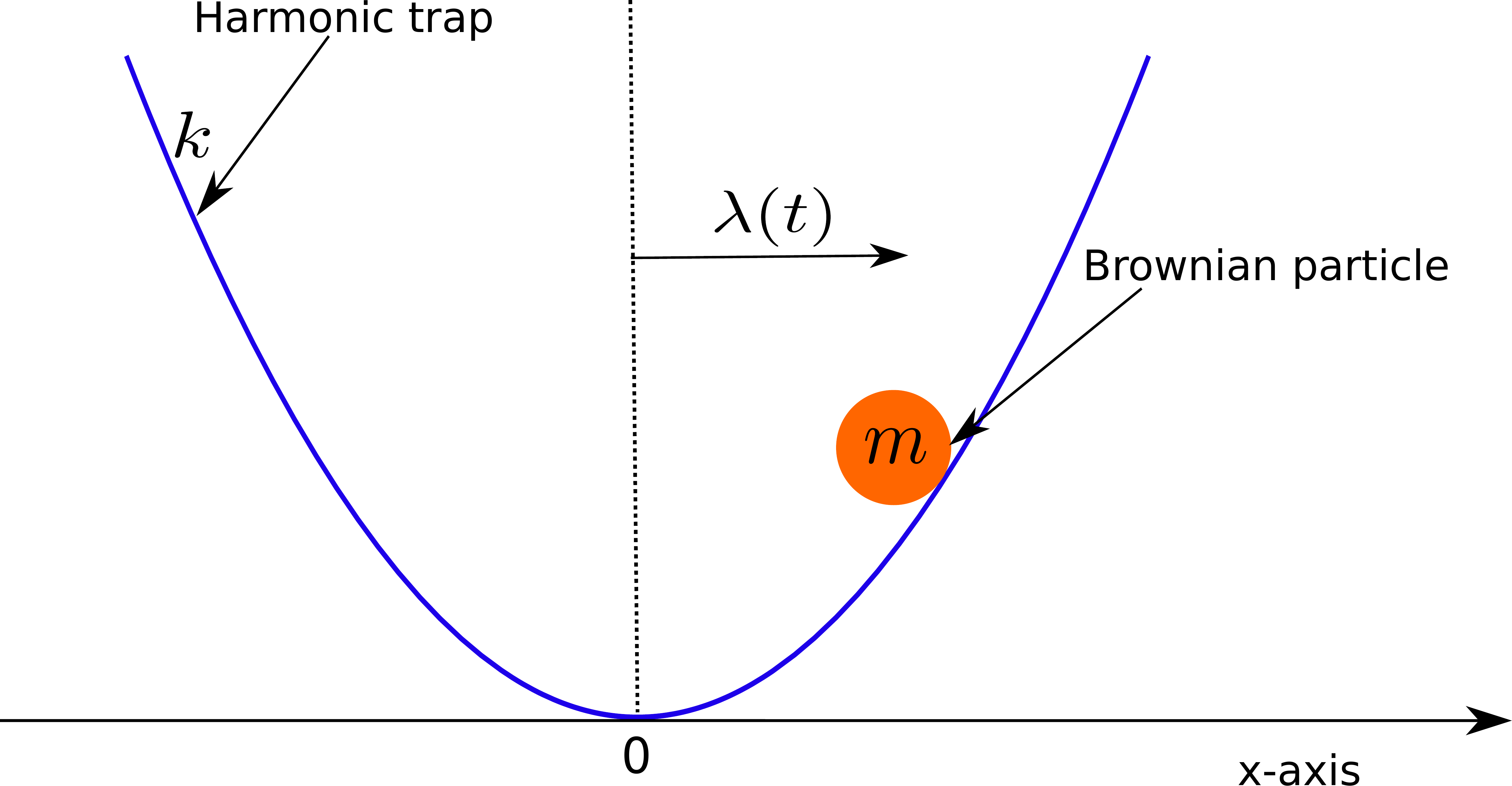}
    \caption{A Brownian particle of mass $m$ is confined in a harmonic trap of stiffness constant $k$. The whole system is immersed in a heat bath (not shown) at temperature $T$.  The vertical dashed line indicates the center of the harmonic trap. The system is driven out of equilibrium by moving the center of the trap along $x$-axis using a protocol $\lambda(t)$ for $t\geq0$.}
    \label{scheme}
  \end{center}
\end{figure}

\section{Model}
\label{sec-model}
Consider a Brownian particle of mass $m$ confined in a harmonic trap of stiffness $k$. The whole setup is in contact with a heat bath at temperature $T$. The center of the confinement  is moved under a protocol $\lambda(t)$ for $t\geq0$. The schematic diagram of the system is shown in \fref{scheme}.  The following underdamped equations describe the dynamics of the system:
\begin{align}
&\dot x=v,\\
&m\dot v=-k[x-\lambda(t)]-\gamma v+\sqrt{2 D \gamma^2}\xi(t)\label{v-eqn},
\end{align} 
where the dot represents a derivative with respect to time, $x$ and $v$, respectively, are the position and velocity of the Brownian particle, $\gamma$ is the dissipation constant, and  $D=\kappa_BT/\gamma$ is the diffusion constant ($\kappa_B$ is the Boltzmann's constant and $T$ is the temperature). In \eref{v-eqn}, $\xi(t)$ is a Gaussian white noise with average $\langle\xi(t)\rangle=0$ and covariance $\langle\xi(t_1)\xi(t_2)\rangle=\delta(t_1-t_2)$. The above equation can be rewritten in the matrix form as
\begin{align}
\dfrac{dU}{dt}=-A U+\dfrac{1}{t_\gamma t_k}F(t)+\sqrt{\dfrac{2 D}{t_\gamma^2}}\eta(t),
\label{mat-eqn}
\end{align}
where $t_\gamma=m/\gamma$, $t_k=\gamma/k$, $U=(x,v)^\top$, $F(t)=(0,\lambda(t))^\top$, $\eta(t)=(0,\xi(t))^\top$, and 
\[ A=
\begin{pmatrix}
  0      && -1 \\
  &&\\
  \dfrac{1}{t_\gamma t_k} && \dfrac{1}{t_{\gamma}}
\end{pmatrix}.
\]
Note that the symbol $\top$ indicates the transpose of a matrix.

Initially for time $t\leq0$, the trap is kept stationary, i.e., $\lambda(t)=0$. Therefore, the system follows the equilibrium Gibbs-Boltzmann distribution:
\begin{align}
P_{\text{eq}}(U)=\dfrac{1}{\sqrt{(2\pi)^2 \det{\Sigma}}}~ \exp\bigg[-\dfrac{1}{2} U^\top \Sigma^{-1} U\bigg],
\label{eq-dist-U}
\end{align}
where the correlation matrix $\Sigma$ is given by
\[ \Sigma=
\begin{pmatrix}
  Dt_k      && 0 \\
  &&\\
  0 && \dfrac{D}{t_\gamma}
\end{pmatrix}.
\]
At $t=0$, the protocol is being switched on. Thus, the formal solution of \eref{mat-eqn}  is
\begin{align}
U(t)&=G(t) U_0+\int_0^t dt_1\  G(t-t_1)\bigg[\dfrac{1}{t_\gamma t_k}F(t_1)+\sqrt{\dfrac{2 D}{t_\gamma^2}}\eta(t_1)\bigg],
\label{U-time-t}
\end{align} 
where $G(t)=e^{-A t}$ and its matrix elements $G_{ij}(t)=[G(t)]_{ij}$ are \footnote{These formulas can be easily derived by noticing that the matrix $B\equiv A-I/(2t_{\gamma})$ is such that $B^{2n}= I[1/(2t_{\gamma})^{2} -1/(t_{\gamma} t_k)]^n$, where $I$ is the identity matrix, and $n$ is a positive integer.} 
\begin{align*}
G_{11}(t)&=e^{-\frac{t}{2 t_\gamma}} \bigg[\frac{\sqrt{t_k} \sinh \bigg(\frac{t \sqrt{t_k-4 t_\gamma}}{2 t_\gamma \sqrt{t_k}}\bigg)}{\sqrt{t_k-4 t_\gamma}}+\cosh \bigg(\frac{t \sqrt{t_k-4 t_\gamma}}{2 t_\gamma \sqrt{t_k}}\bigg)\bigg],\\
G_{12}(t)&=\frac{2 t_\gamma t_k e^{-\frac{t}{2 t_\gamma}} \sinh \left(\frac{t \sqrt{t_k (t_k-4 t_\gamma)}}{2 t_\gamma t_k}\right)}{\sqrt{t_k (t_k-4 t_\gamma)}},\\
G_{21}(t)&=-\frac{2 e^{-\frac{t}{2 t_\gamma}} \sinh \left(\frac{t \sqrt{t_k (t_k-4 t_\gamma)}}{2 t_\gamma t_k}\right)}{\sqrt{t_k (t_k-4 t_\gamma)}},\\
G_{22}(t)&=e^{-\frac{t}{2 t_\gamma}} \bigg[\cosh \bigg(\frac{t \sqrt{t_k-4 t_\gamma}}{2 t_\gamma \sqrt{t_k}}\bigg)-\frac{\sqrt{t_k} \sinh \bigg(\frac{t \sqrt{t_k-4 t_\gamma}}{2 t_\gamma \sqrt{t_k}}\bigg)}{\sqrt{t_k-4 t_\gamma}}\bigg],
\end{align*}
and $U_0=U(0)$.

Since $U(t)$ is linear in the Gaussian thermal white noise, the mean and correlation are sufficient to obtain its probability density function. Averaging over both initial condition $U_0$ with respect to $P_{\text{eq}}(U_0)$ [see \eref{eq-dist-U}] and Gaussian thermal white noise $\xi(t)$, we obtain mean and correlation of $U(t)$ (see \aref{jpdf}) as
\begin{align}
&\overline{\langle U(t)\rangle}=\dfrac{1}{t_\gamma t_k}\int_0^t dt_1\  G(t-t_1) F(t_1),\\
&\overline{\langle U(t) U^\top(t)\rangle}-\overline{\langle U(t)\rangle}\ \overline{\langle U^\top(t)\rangle}=\Sigma\label{var-U},
\end{align}
where angular brackets represent the average over the Gaussian thermal white noise and the overhead bar indicates the average over the initial condition $U_0$ with respect to $P_{\text{eq}}(U_0)$. \textcolor{black}{Notice that the variance of $U$ does not change with the time [see \erefss{eq-dist-U} and \erefs{var-U}] when one drags the center of the trap using a protocol $\lambda(t)$.}
Therefore, the probability density function of $U$ at time $t$ is
\begin{align}
P(U,t)&=\dfrac{1}{\sqrt{(2\pi)^2 \det{\Sigma}}}~\exp\bigg[-\dfrac{1}{2} \big[U-\overline{\langle U(t)\rangle}\big]^\top\  \Sigma^{-1}\  \big[U-\overline{\langle U(t)\rangle}\big]\bigg].
\label{dist-U}
\end{align}

In this paper, we focus on two observables: the position of the particle $x(t)$ (an even variable with respect to time) and the current $\phi(t)$ (an odd variable with respect to time) at time $t$ defined as
\begin{align}
\phi(t):=\int_0^t dt'\  \dot x(t')=x(t)-x(0),
\end{align}
and our aim is to obtain the thermodynamics uncertainty relations corresponding to them, i.e., $\text{Var}[x(t)]/\overline{\langle x(t)\rangle}^2$ and $\text{Var}[\phi(t)]/\overline{\langle \phi(t)\rangle}^2$, where $\text{Var}[f(t)]:=\overline{\langle f(t)^2\rangle}-\overline{\langle f(t)\rangle}^2$ is the variance of a function $f(t)$.

\section{Thermodynamic uncertainty relations}
\label{sec-tur}
It is evident that the system in thermal equilibrium does not generate entropy. Therefore, the total entropy production $\Delta s_{tot}=0$, where $\Delta s_{tot}$ along a stochastic trajectory is defined as \cite{Seifert-entropy}
\begin{align}
\Delta s_{tot}&=\Delta s_{med}+\Delta s_{sys}\nonumber\\
&=-\dfrac{Q}{T}-\ln P(U,t)+\ln P_{\text{eq}}(U_0).
\label{s-tot-UD}
\end{align}
On the right-hand side, the first term and last two terms, respectively, are the medium and system entropy production from time $0$ to $t$. In the above equation, $Q$ is the amount of the heat absorbed by the system from the heat reservoir.

When a system is driven out of equilibrium using a non-equilibrium protocol, the system generates entropy, moreover, this entropy production [see \eref{s-tot-UD}] is a stochastic quantity, i.e., its value changes from one realization to another. The total entropy production satisfies a well-known identity called the \emph{integral fluctuation theorem} \cite{Seifert-entropy}:
\begin{align}
\overline{\langle e^{-\Delta s_{tot}}\rangle}=1, \label{IFT}
\end{align}
where angular brackets represents the average over realizations of the experiment and overhead bar indicates the average over the initial condition. Using the Jensen's inequality, i.e.,  $\overline{\langle e^{-\Delta s_{tot}}\rangle}\geq e^{\overline{\langle -\Delta s_{tot} \rangle}}$, in Eq. \eqref{IFT}, we can show that
\begin{align}
\overline{\langle\Delta s_{tot}\rangle}\geq0,
\label{ii-law}
\end{align}
which is the second law of thermodynamics. In our case, using $P(U,t)$ and $P_{\text{eq}}(U_0)$ in $\Delta s_{sys}$, and averaging over both initial $U_0$ and final variables $U$, one can show that $\overline{\langle\Delta s_{sys}\rangle}=0$. Therefore, \eref{ii-law} modifies to
\begin{align}
\overline{\langle\Delta s_{med}\rangle}\geq0.
\label{m-ii-law}
\end{align}

In the following, we identify $\Delta s_{med}=-Q/T$ along a single stochastic trajectory. 
Multiplying $v$ on both sides of \eref{v-eqn}, rearranging terms, and integrating over time from $0$ to $t$, yields the \emph{first law of thermodynamics} \cite{Sekimoto}:
\begin{align}
\Delta E=W+Q\label{f-l-U},
\end{align}
where we identify terms
\begin{align}
\Delta E&=\int_0^t dt'\  \dfrac{d}{dt'} \bigg(\dfrac{1}{2}m v(t')^2+\dfrac{1}{2}k[x(t')-\lambda(t')]^2\bigg)\nonumber\\
&=\dfrac{1}{2}m (v^2-v_0^2)+\dfrac{1}{2}k\big[\{x-\lambda(t)\}^2-x_0^2\big],\\
W&=k\int_0^t dt' [\lambda(t')-x(t')]\dot{\lambda}(t'),\\
Q&=\int_0^t dt' [\sqrt{2 D\gamma^2 }\xi(t')-\gamma v(t')]v(t'),\label{Q-eqn}
\end{align}
as change in the internal energy ($\Delta E$), work done ($W$) on a Brownian particle by moving the harmonic trap, and heat ($Q$) absorbed by a Brownian particle from the heat reservoir, respectively. Notice that the integral  in \eref{Q-eqn} follows the Stratonovich rule of integration \cite{Sekimoto}.
Using  \erefss{s-tot-UD}, \erefs{m-ii-law}, and \erefs{f-l-U}, and averaging over all realizations, we find that
\begin{align}
\overline{\langle \mathcal{W}\rangle}\geq \dfrac{1}{T}\overline{\langle \Delta E\rangle},
\label{w-e}
\end{align}
where $\mathcal{W}=\frac{1}{T} W$ is the dimensionless work done measured in the units of the temperature $T$ of the bath. Notice that we have set Boltzmann's constant $\kappa_B$ equal to one. 
{\color{black}In the above \eref{w-e}, the right-hand side is 
\begin{align}
\dfrac{1}{T}\overline{\langle \Delta E\rangle}=\dfrac{t_\gamma}{2D} \overline{\langle v\rangle }^2+\dfrac{1}{2Dt_k} [\overline{\langle x \rangle}-\lambda(t) ]^2.
\label{part}
\end{align}
}

Using \eref{part} in \eref{w-e}, we obtain the following inequality 
\begin{align}
\dfrac{\mathcal{R}\  \text{Var}[x(t)]}{\overline{\langle x\rangle}^2}\geq \dfrac{1}{2} \bigg[1-\dfrac{\lambda(t)}{\overline{\langle x\rangle}}\bigg]^2, 
\label{UR-x}
\end{align}
where $\text{Var}[x(t)]= Dt_k$ and
\begin{align}
\mathcal{R}&=\overline{\langle \mathcal{W}\rangle}-\dfrac{t_\gamma}{2 D}\overline{\langle v\rangle}^2, \\
\overline{\langle x\rangle}&=\frac{1}{t_\gamma t_k}\int_0^t dt_1\  G_{12}(t-t_1)\lambda(t_1),\\
\overline{\langle v\rangle}&=\dfrac{1}{t_\gamma t_k}\int_0^t dt_1\  G_{22}(t-t_1)\lambda(t_1)=\dfrac{1}{t_\gamma t_k}\int_0^t dt_1\  G_{12}(t-t_1)\dfrac{\partial\lambda(t_1)}{\partial t_1}. \label{v-bar-eqn}
\end{align}
In the above equation \eqref{v-bar-eqn}, we have used that $G_{22}(t)=\frac{\partial G_{12}(t)}{\partial t}$ and integration by parts.
The relation \erefs{UR-x} is the thermodynamic uncertainty relation for the position variable. An alternative approach to obtain \eref{UR-x} is as follows. 
In the joint probability density function $\mathcal{P}(x,v,\mathcal{W},t)$ (see \aref{jpdf}), the covariance matrix $\tilde{\Sigma}$ is positive semi-definite. Therefore, its determinant is 
\begin{align} 
\overline{\langle (\delta x)^2\rangle}\ \overline{\langle (\delta v)^2\rangle}\ \overline{\langle (\delta \mathcal{W})^2\rangle}-\overline{\langle (\delta x)^2\rangle}\ \overline{\langle \delta \mathcal{W} \delta v\rangle}^2- \overline{\langle \delta \mathcal{W} \delta x\rangle}^2\overline{\langle (\delta v)^2\rangle}\geq 0.
\end{align} 
Using \erefss{var-U}, \erefs{wx-f}, \erefs{wv-f}, and \erefs{varw}, we can prove \eref{UR-x}.

The average of the observable current $\phi(t)$ is equal to that of $x(t)$\footnote{The quantity $\phi(t)=x(t)-x_0$. Averaging over the initial condition given by $P_{eq}(U_0)$, gives $\overline{\phi}=\overline{x(t)}$. Finally averaging over the Gaussian white noise yields $\overline{\langle \phi(t) \rangle}=\overline{\langle x(t)\rangle}$.}, whereas the variance of $\phi$ is  
\begin{align}
\text{Var}[\phi(t)]=2 D t_k[1-G_{11}(t)].
\end{align}
Thus, the similar thermodynamic uncertainty relation for $\phi(t)$ can be obtained as
\begin{align}
\dfrac{\mathcal{R}\  \text{Var}[\phi(t)]}{\overline{\langle \phi\rangle}^2}&\geq[1-G_{11}(t)]\bigg[1-\dfrac{\lambda(t)}{\overline{\langle \phi\rangle}}\bigg]^2.
\label{UR-phi}
\end{align}

In the following, we obtain the thermodynamic uncertainty relations for position and current variables in the overdamped limit, i.e., $t_\gamma\to0$. In this limit, the mean position and mean velocity of the particle are obtained as 
\begin{align}
\overline{\langle x\rangle}&=\frac{1}{t_k}\int_0^t dt_1\  e^{-(t-t_1)/t_k}\lambda(t_1),\label{x-mean-od}\\
\overline{\langle v\rangle}&=\dfrac{1}{ t_k}\int_0^t dt_1\  e^{-(t-t_1)/t_k}\dfrac{\partial\lambda(t_1)}{\partial t_1}.
\end{align}
Therefore, we find that $\mathcal{R}=\overline{\langle \mathcal{W}\rangle}$ in the overdamped limit. Moreover, in the same limit, the variance of $\phi$ is $2 Dt_k(1-e^{-t/t_k})$.

The thermodynamic uncertainty relations for position and current variables in the overdamped limit ($t_\gamma\to0$)  can be obtained as
\begin{align}
\dfrac{\overline{\langle\mathcal{W}\rangle}\  \text{Var}[x(t)]}{\overline{\langle x\rangle}^2}&\geq\dfrac{1}{2}\bigg[1-\dfrac{\lambda(t)}{\overline{\langle x\rangle}}\bigg]^2,\label{ur-x-od}\\
\dfrac{\overline{\langle\mathcal{W}\rangle}\  \text{Var}[\phi(t)]}{\overline{\langle \phi\rangle}^2}&\geq(1-e^{-\frac{t}{t_k}})\bigg[1-\dfrac{\lambda(t)}{\overline{\langle \phi\rangle}}\bigg]^2\label{ur-phi-od}.
\end{align}
Note that the right-hand side of the above equations [\erefss{ur-x-od} and \erefs{ur-phi-od}] depends only on the external protocol acting on the system through \eref{x-mean-od}, and it depends on the choice of the system studied.
When a protocol $\lambda(t)$ vanishes at a final time of the observation, the right-hand side of the uncertainty relations simplifies to $1/2$ and $(1-e^{-t/t_k})$ in \erefss{ur-x-od} and \erefs{ur-phi-od}, respectively. In such cases, the cost function is given by $\overline{\langle\mathcal{W}\rangle}$. Moreover, when we substitute $\tau=\mathcal{T}/t_k$ (dimensionless time), where $t=\mathcal{T}$ is the observation time such that $\lambda(\mathcal{T})=0$ in \eref{ur-phi-od}, and using the large time limit $\tau\gg1$, we see that the right-hand side approaches to unity instead of 2 \cite{general}.

\begin{figure*}[!h]
  \begin{tabular}{ccc}
    \includegraphics[width=5cm]{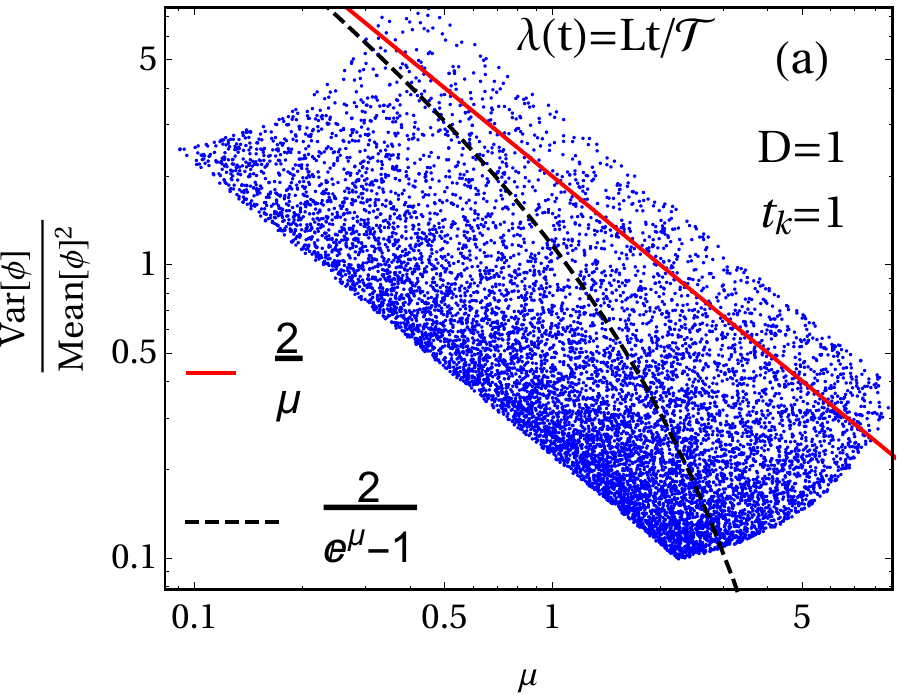}~~~~~~~
    \includegraphics[width=5cm]{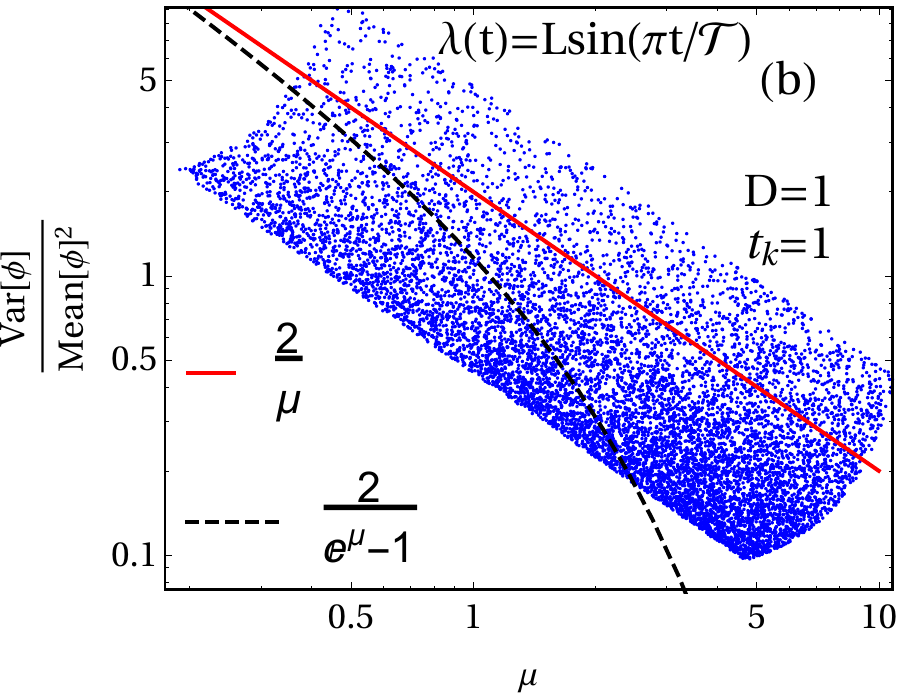}~~~~~~~~
    \includegraphics[width=5cm]{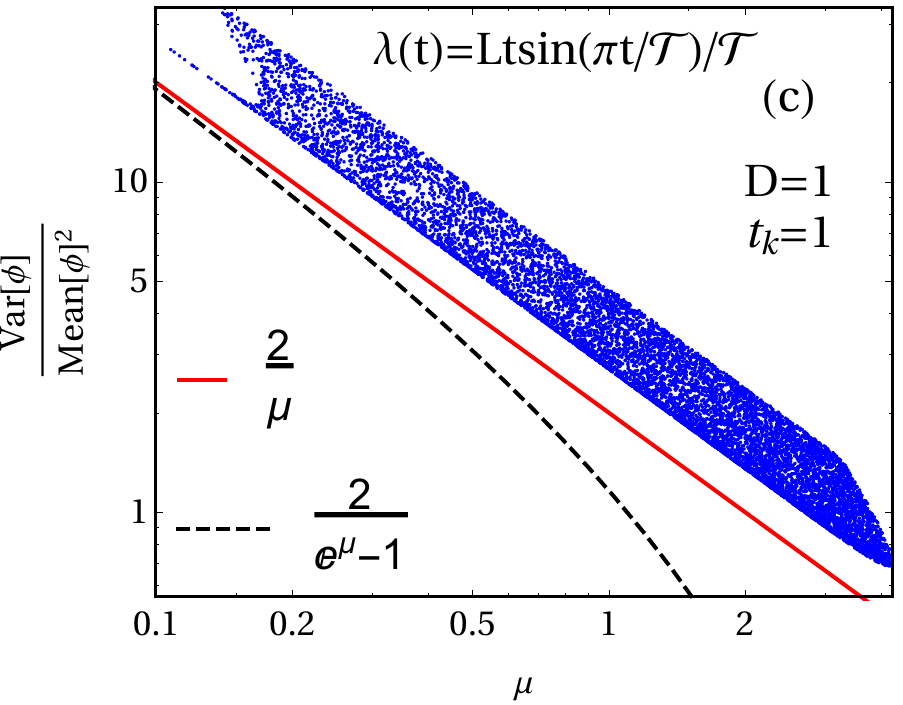}
    \end{tabular}
    \caption{{\color{black}$\frac{\rm{Var}[\phi(t)]}{\overline{\langle \phi\rangle}^2}$ with respect to $\mu=\overline{\langle\mathcal{W}\rangle}$ for protocols (a) $\lambda(t)=\frac{L t}{\mathcal{T}}$, (b) $\lambda(t)=L{\sin\big(\frac{\pi t}{\mathcal{T}}\big)}$, and $\lambda(t)=\frac{Lt}{\mathcal{T}}{\sin\big(\frac{\pi t}{\mathcal{T}}\big)}$  at a time (a) $t=\mathcal{T}$, (b) $t=\mathcal{T}/2$, and (c) $t=\mathcal{T}$, where $L$ and $\mathcal{T}$ are positive constants having the dimension of length and time, respectively. Red solid and black dashed lines, respectively, correspond to $\frac{2}{\mu}$ and $\frac{2}{e^\mu-1}$ \cite{general}. Blue dots are obtained using uniformly distributed random numbers $L\in[1,5]$, and $\tau\in[1,10]$, where $\tau=\mathcal{T}/t_k$ is the dimensionless time. Here, the plots are given for fix parameters $D=1$, and $t_k=1$. We see that the quantity $\phi$ obeys the bound given in \cite{general} only for those protocol which vanishes at the final time. } }
    \label{fig:sim}
\end{figure*}
\begin{figure*}
  \begin{tabular}{ccc}
    \includegraphics[width=5cm]{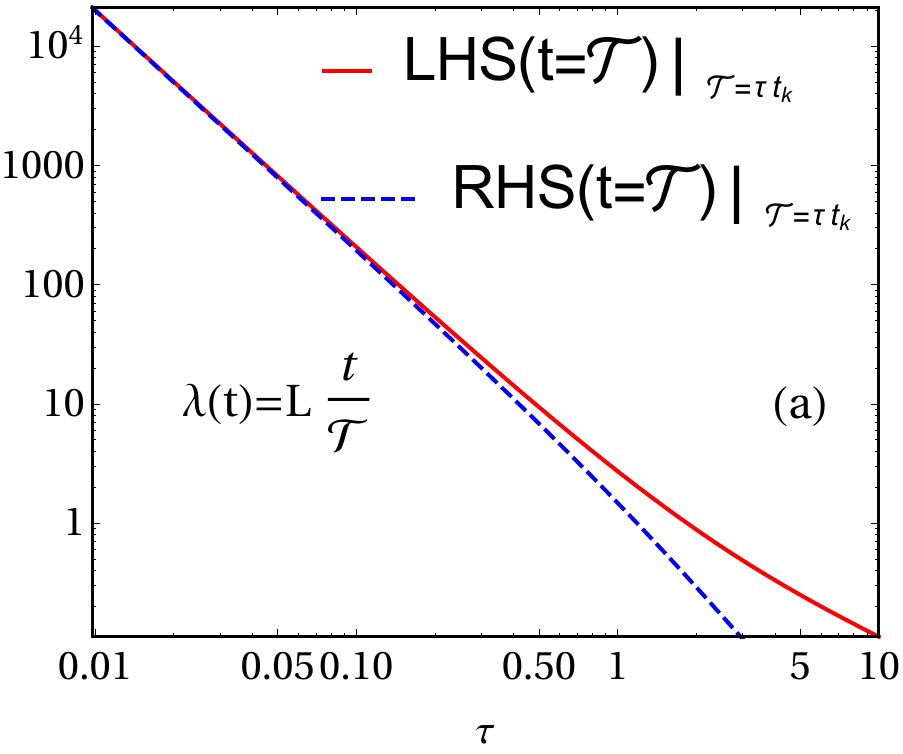}~~~~~~~
    \includegraphics[width=5cm]{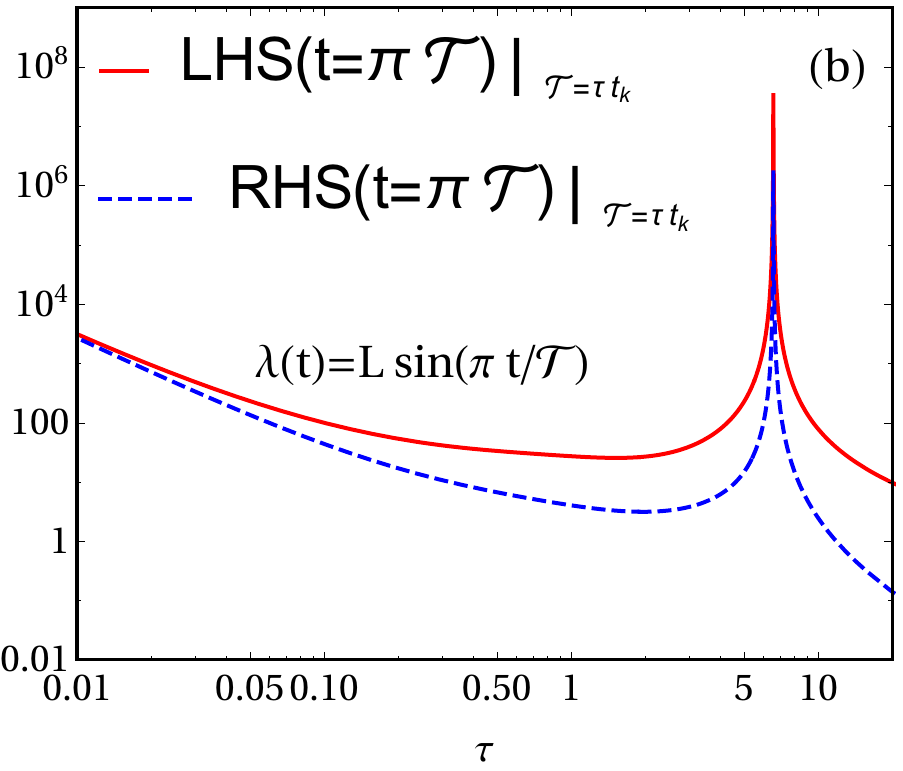}~~~~~~~~
    \includegraphics[width=5cm]{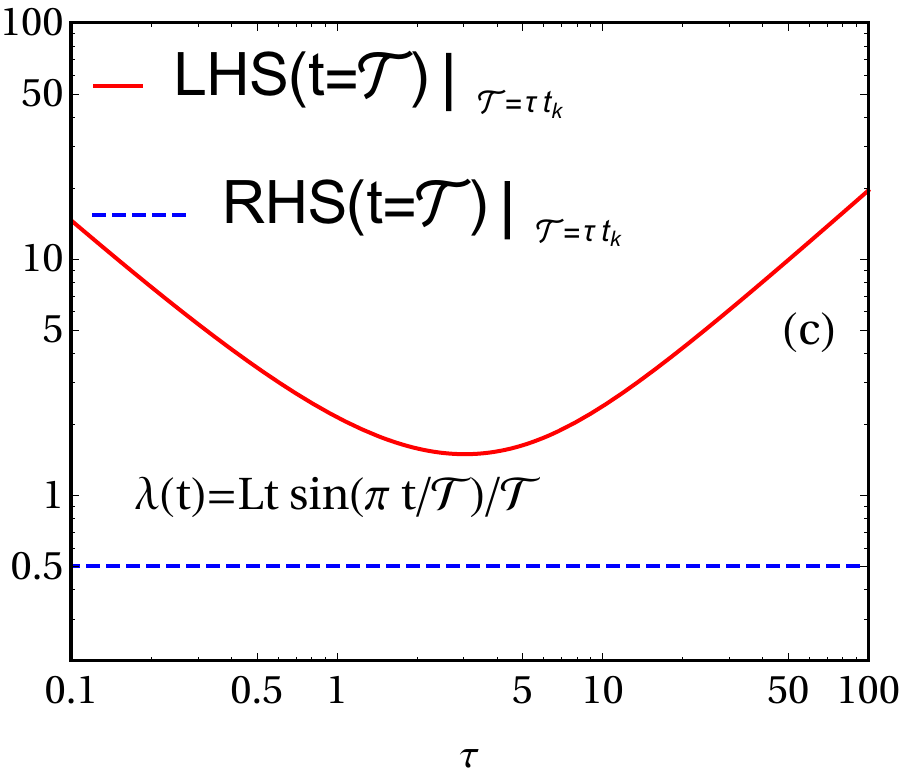}
  \end{tabular}
    \caption{(LHS) Left- (red solid line) and (RHS) right-hand sides (blue dashed line) of \eref{ur-x-od} are plotted against $\tau$ for three protocols: (a) $\lambda(t)=L \frac{t}{\mathcal{T}}$, (b) $\lambda(t)=L \sin\big(\pi\frac{t}{\mathcal{T}}\big)$, (c) $\lambda(t)=L\frac{t}{\mathcal{T}} \sin\big(\pi\frac{t}{\mathcal{T}}\big)$, where $L$ and $\mathcal{T}$ are positive constants having the dimension of length and time, respectively, and $\tau=\mathcal{T}/t_k$ is the dimensionless time. We can see that the red solid line always stays above the blue dashed line irrespective of the choice of the external protocol. }
    \label{fig:ur-x-od}
  \end{figure*}
\begin{figure*}
  \begin{tabular}{ccc}
    \includegraphics[width=5cm]{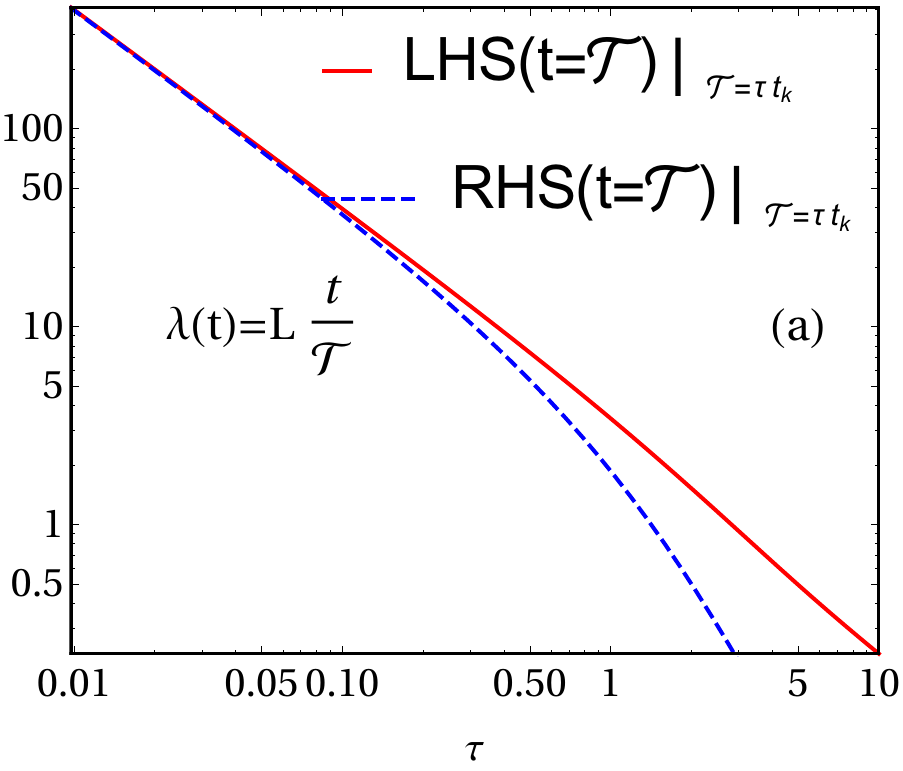}~~~~~~~
    \includegraphics[width=5cm]{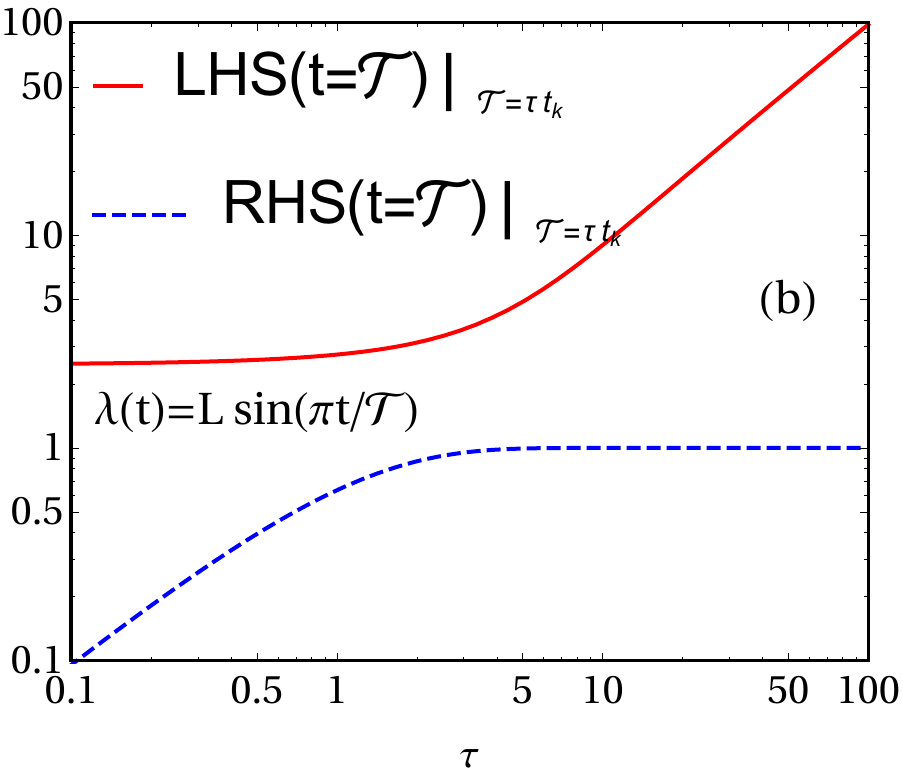}~~~~~~~
    \includegraphics[width=5cm]{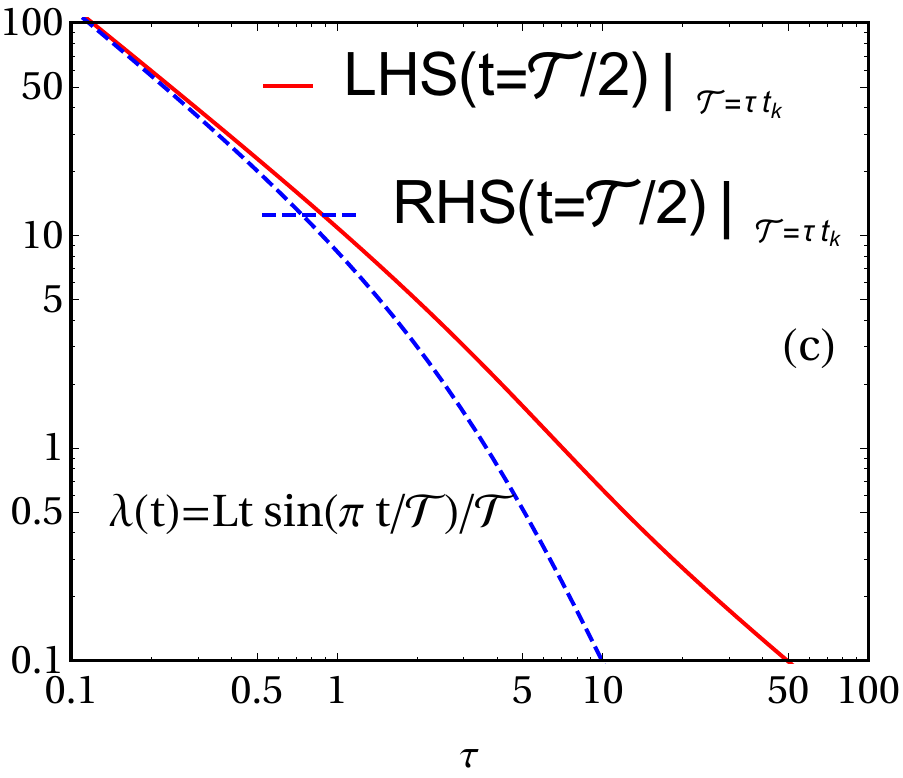}
    \end{tabular}
    \caption{(LHS) Left- (red solid line) and (RHS) right-hand sides (blue dashed line) of the thermodynamic uncertainty relation \eref{ur-phi-od} for $\phi$ are plotted against $\tau$ for three different protocols: (a) $\lambda(t)=L \frac{t}{\mathcal{T}}$, (b) $\lambda(t)=L \sin\big(\pi\frac{t}{\mathcal{T}}\big)$, (c) $\lambda(t)=L\frac{t}{\mathcal{T}} \sin\big(\pi\frac{t}{\mathcal{T}}\big)$, where $L$ and $\mathcal{T}$ are positive constants having the dimension of length and time, respectively, and $\tau=\mathcal{T}/t_k$ is the dimensionless time. The red solid line always stays above the blue dashed line irrespective of the choice of the external protocol. }
    \label{fig:ur-phi-od}
\end{figure*}
  
{\color{black}
In \fref{fig:sim}, we plot $\text{Var}[\phi]/\text{Mean}[\phi]^2$ with respect to the average work done ${\overline{\langle \mathcal{W} \rangle }}$ for three different protocols: (a) $\lambda(t)=L \frac{t}{\mathcal{T}}$, (b) $\lambda(t)=L \sin\big(\pi\frac{t}{\mathcal{T}}\big)$, (c) $\lambda(t)=L\frac{t}{\mathcal{T}} \sin\big(\pi\frac{t}{\mathcal{T}}\big)$, where $L$ and $\mathcal{T}$ are positive constants having the dimension of length and time, respectively. The red solid and black dashed lines, respectively, are given by functions $2/\overline{\langle \mathcal{W}\rangle}$, and $2/(e^{\overline{\langle \mathcal{W} \rangle }}-1)$ \cite{general}.  The blue dots for $\text{Var}[\phi]/\text{Mean}[\phi]^2$ are obtained for uniformly distributed random values of $L\in[1,5]$ and $\tau\in[1,10]$ at fixed parameters $D=1$ and $t_k=1$. It is clear from the figure that when the protocol does not vanish at the final time [for example, \frefss{fig:sim}(a) and (b)], the uncertainty ratio does not obey the bounds  $2/\overline{\langle \mathcal{W}\rangle}$ and $2/(e^{\overline{\langle \mathcal{W} \rangle }}-1)$ (which is true since the protocol is not time-symmetric), whereas as one considers the protocol (b) (time-symmetric) such that it vanishes at the final time, the relation hold (not shown) as predicted in \cite{general} because the joint distribution of $\phi$ and $\mathcal{W}$ obeys the fluctuation theorem.  However, we \textit{surprisingly} found that the bound given in \cite{general} holds [see \fref{fig:sim}(c)] even for the protocol $L\frac{t}{\mathcal{T}} \sin(\pi \frac{t}{\mathcal{T}})$ (time asymmetric protocol) as all the blue points are above the curve $2/\overline{\langle \mathcal{W}\rangle}$ only when final $t$ is such that $\lambda(t)=0$. Thus, we find that the previously observed bounds \cite{general} may not even hold  for all range of parameters for such a simple setup used in this paper.   

In contrast to the previous observation, in the following, we test the bounds [\erefss{ur-x-od} and \erefs{ur-phi-od}] presented in this paper.
In \frefss{fig:ur-x-od} and \frefs{fig:ur-phi-od}, we plot the left- (red solid line) and right-hand sides (blue dashed line) of thermodynamic uncertainty relations \erefs{ur-x-od} and \erefs{ur-phi-od}, respectively,  against $\tau$ for the above given three protocols. 
Three possibilities can be seen depending on either at (P1) the shorter time [see \frefss{fig:ur-x-od}(a), \frefs{fig:ur-phi-od}(a), and \frefs{fig:ur-phi-od}(c)], (P2) the intermediate time [see \frefss{fig:ur-x-od}(c) and \frefs{fig:ur-phi-od}(b)] or (P3) at both shorter and intermediate times [see \fref{fig:ur-x-od}(b)], the red solid curve gets closer to the blue dashed one as a function of the observation time $\tau$. These possibilities depend on both protocol $\lambda(t)$ and observation time $\tau$. We study two different types of protocols depending on whether (A) the external protocol return to its initial value $\lambda(0)=0$ or (B) the center is displaced by a finite distance in a time interval $\tau$. For protocol (A), the mean work done tends to zero in the limit $\tau \to 0$ whereas it remains non-zero for protocol (B). This is because the protocol (A) performs work of an order of $\tau$ on the Brownian particle for small $\tau$ whereas in the case of protocol (B), the work done remains non-zero in this limit since it corresponds to a  fast non-equilibrium process to move the center of the trap to a non-zero distance in a very short interval of time [see \eref{s-w}].  Further, we notice that mean work done approaches to zero in the limit $\tau\to \infty$ (quasi-static process) [see \eref{l-w}], and the mean distance tends to zero in the limit $\tau\to 0$ [see \eref{s-x}] and approaches to $\lambda(\mathcal{T})$ in the limit of large-$\tau$ [see \eref{l-x}] for both types of protocols [(A) and (B)]. With these observations, we can discuss various possibilities (P1)--(P3). Let us first consider the possibility (P1). At the short time, both left- and right-hand sides of \eref{ur-x-od} diverge and approach to each other whereas they vanish at a large time with different orders in $\tau$. Similar behavior can also be seen in \eref{ur-phi-od}. In the case of possibility (P2), both mean position and mean work done tend to zero at both small and large-time $\tau$, and attain maximum values at the intermediate time. Moreover, the ratio $\overline{\langle\mathcal{W}\rangle}\text{Var}[x]/\overline{\langle x\rangle}^2$ diverges at both small and large times and has a minimum at a particular $\tau$ [see \fref{fig:ur-x-od}(c)] due to non-zero values of $\overline{\langle x\rangle}$ and $\overline{\langle \mathcal{W}\rangle}$.   Since the variance of current $\phi$ in \erefss{ur-phi-od} is of an order $\tau$ in the limit $\tau\to 0$ and becomes constant at large time, the quantity $\overline{\langle\mathcal{W}\rangle}\text{Var}[\phi]/\overline{\langle \phi\rangle}^2$ initially remains constant and then diverges in the limit $\tau\to \infty$. For possibility (P3) [\fref{fig:ur-x-od}(b)], the short time behaviour is similar to (P1). Moreover, we see that the  red solid and blue dashed curves diverge at a particular $\tau$. This is because at that $\tau$ the mean position becomes zero  whereas the mean work done attains a non-zero value.   

It is clear from \frefss{fig:ur-x-od} and \frefs{fig:ur-phi-od} that both bounds are satisfied for all $\tau$ irrespective of the choice of the external protocol as compared to \fref{fig:sim}. Therefore, the thermodynamic uncertainty relations given in \erefss{ur-x-od} and \erefs{ur-phi-od} obtained by invoking the second law of thermodynamic, do not require several constraints such as validity of fluctuation theorem, time-symmetric nature of the protocol, measurement of the observable in the time-reversed trajectory, etc..

\section{Summary}
\label{summ}
We have considered a Brownian particle confined in a harmonic trap in one dimension in the underdamped regime. The system is moved away from the equilibrium by moving the center of harmonic trap using an arbitrary protocol $\lambda(t)$ along the $x$-axis. Using the second law of thermodynamics (i.e., $\overline{\langle \Delta s_{tot} \rangle}\geq0$, where $\overline{\langle \Delta s_{tot} \rangle}$ is the average total entropy production), we obtained the thermodynamic uncertainty relation for both  position (even variable) and current (odd variable)} at time $t$. Further, we obtained these relations in the overdamped limit ($t_\gamma\to0$). The generalization of this result for a general current in a generic setup is still not clear, and it would be great to understand this problem in future.

As a final remark, our setup can be realized in an experiment (for example, similar setup is already used in experiments shown in \cite{Wang-2,Wang-1}), and it would be interesting to test our theoretical predictions experimentally.

\section*{Acknowledgements}D. Gupta and A. Maritan  acknowledge the support from University of Padova  through  `Excellence Project 2018'' of the Cariparo foundation.
\section*{Author contributions}D. Gupta and A. Maritan designed research, D. Gupta performed research, and both the authors discussed the results and wrote the paper.

\appendix

\section{Joint probability density function: $\mathcal{P}(x,v,\mathcal{W},t)$}
\label{jpdf}
In this section, we compute the joint probability density function: $\mathcal{P}(x,v,\mathcal{W},t)$. Since all the variables: $x$, $v$ and $\mathcal{W}$ are Gaussian random variables, their joint probability density function will also be Gaussian. To obtain it, we first compute the the auto-correlation function of $U$ using \eref{U-time-t} at two different times (where $t_1\leq t_2$): 
\begin{align}
&\overline{\langle U(t_1) U^\top(t_1)\rangle}-\overline{\langle U(t_2)\rangle}\ \overline{\langle U^\top(t_2)\rangle}=G(t_1) \Sigma G^\top(t_2)+\dfrac{2D}{t_\gamma^2}\int_0^{t_1} dt^\prime_1\  G(t_1-t_1^\prime) M G^\top(t_2-t_1^\prime),
\label{auto}
\end{align}
where $M_{ij}=\delta_{i,2}\delta_{i,j}$. 

Note that $\dfrac{2D}{t_\gamma^2}M=A\Sigma+\Sigma A^\top$,  $\dfrac{d}{dt_1^\prime}G(t_1-t_1^\prime)=G(t_1-t_1^\prime)A$, and $\dfrac{d}{dt_1^\prime}G^\top(t_1-t_1^\prime)=A^\top G^\top(t_1-t_1^\prime)$. Therefore, we find $G(t_1-t_1^\prime) M G^\top(t_2-t_1^\prime)=\dfrac{d}{dt_1^\prime}\bigg[G(t_1-t_1^\prime) \Sigma G^\top(t_2-t_1^\prime)\bigg]$,

We perform the integral in \eref{auto}, and rewrite it as
\begin{align}
\overline{\langle U(t_1) U^\top(t_1)\rangle}-\overline{\langle U(t_2)\rangle}\ \overline{\langle U^\top(t_2)\rangle}&=G(0)\Sigma G^\top(t_2-t_1), 
\label{auto-2}
\end{align}
where $G(0)=\mathcal{I}_{2\times 2}$ is an identity matrix.  Similarly, for $t_1\geq t_2$, we get 
\begin{align}
\overline{\langle U(t_1) U^\top(t_1)\rangle}-\overline{\langle U(t_2)\rangle}\ \overline{\langle U^\top(t_2)\rangle}&= G(t_1-t_2)\Sigma G^\top(0).
\label{auto-3}
\end{align}
Setting $t_1=t_2=t$, we can obtain \eref{var-U}.

Now our aim is to compute the other correlations at time $t$ such as $\overline{\langle \delta \mathcal{W} \delta x\rangle}$, $\overline{\langle \delta \mathcal{W} \delta v\rangle}$, and $\overline{\langle (\delta \mathcal{W})^2\rangle}$,   where $\delta f=f-\overline{\langle f\rangle }$. Therefore,
\begin{align}
\overline{\langle \delta \mathcal{W} \delta x\rangle}&=-\dfrac{1}{Dt_k} \int_0^t dt_1\ \dot{\lambda}(t_1)\   \overline{\langle \delta x(t_1) \delta x(t)\rangle}, \label{wx}\\
\overline{\langle \delta \mathcal{W} \delta v\rangle}&=-\dfrac{1}{Dt_k} \int_0^t dt_1\ \dot{\lambda}(t_1)\   \overline{\langle \delta x(t_1) \delta v(t)\rangle}.\label{wv}
\end{align}

Using \eref{auto-2}, we substitute 
\begin{align}
\overline{\langle \delta x(t_1) \delta x(t)\rangle}&=Dt_k G_{11}(t-t_1),\\
\overline{\langle \delta x(t_1) \delta v(t)\rangle}&=Dt_k G_{21}(t-t_1),
\end{align}
into  \erefss{wx} and \erefs{wv}, respectively. Integrating by parts \erefss{wx} and \erefs{wv}, and using $\dfrac{d}{dt_1}G_{11}(t-t_1)=\dfrac{G_{12}(t-t_1)}{t_\gamma t_k}$, and $\dfrac{d}{dt_1}G_{21}(t-t_1)=\dfrac{G_{22}(t-t_1)}{t_\gamma t_k}$, one can see that
\begin{align}
\overline{\langle \delta \mathcal{W} \delta x\rangle}&=-\lambda(t)+\overline{\langle x \rangle},\label{wx-f}\\
\overline{\langle \delta \mathcal{W} \delta v\rangle}&=\overline{\langle v \rangle}.\label{wv-f}
\end{align}
Similarly, the variance of $\mathcal{W}$ can be computed as 
\begin{align}
\rm{Var}{\mathcal{W}}&=\dfrac{1}{(Dt_k)^2} \int_0^t dt_1\  \dot{\lambda}(t_1) \int_0^t  dt_2\ \dot{\lambda}(t_2) \overline{\langle \delta x(t_1) \delta x(t_2)\rangle}\nonumber\\
&= \dfrac{1}{(Dt_k)^2}\bigg[ \int_0^t dt_1\  \dot{\lambda}(t_1) \int_0^{t_1} dt_2\  \dot{\lambda}(t_2) \overline{\langle \delta x(t_1) \delta x(t_2)\rangle}+\nonumber\\
&~~~~~~~~~+\int_0^t dt_1\  \dot{\lambda}(t_1)\int_{t_1}^t dt_2\  \dot{\lambda}(t_2) \overline{\langle \delta x(t_1) \delta x(t_2)\rangle}\bigg].\nonumber\\
&=\dfrac{2}{(Dt_k)^2} \int_0^t dt_1\  \dot{\lambda}(t_1) \int_0^{t_1}  dt_2\  \dot{\lambda}(t_2) \overline{\langle \delta x(t_1) \delta x(t_2)\rangle}
\label{varw0}
\end{align}
While coming from the second to third equality, we follow two steps: (1) we swap the last two integrals in the second term in the second equality, and (2) interchange the dummy variables $t_1$ and $t_2$, i.e., $t_1\leftrightarrow t_2$.

We use \eref{auto-3} in \eref{varw0}, and then, integrating \erefs{varw0} by parts, we finally obtain
\begin{align}
\rm{Var}{\mathcal{W}}=2\overline{\langle \mathcal{W}\rangle}.\label{varw}
\end{align}
Now using correlations given above one can write the joint probability density function $\mathcal{P}(x,v,\mathcal{W},t)$ at time $t$ starting from the initial distribution $\delta(\mathcal{W})P_{eq}(x_0,v_0)$ at time $t=0$:
\begin{align}
\mathcal{P}(\mathcal{U},t)&=\dfrac{1}{\sqrt{(2\pi)^3\  \det[\tilde{\Sigma}]}} ~\exp\bigg[-\dfrac{1}{2} \big[\mathcal{U}-\overline{\langle \mathcal{U}\rangle}\big]^\top\  \tilde{\Sigma}^{-1}\  \big[\mathcal{U}-\overline{\langle \mathcal{U}\rangle}\big]\bigg],
\end{align}
where $\mathcal{U}=(x,v,\mathcal{W})^\top$, and $\tilde{\Sigma}_{i,j}=\overline{\langle \mathcal{U}_i(t) \mathcal{U}_j(t)\rangle}-\overline{\langle \mathcal{U}_i(t)\rangle}\ \overline{\langle \mathcal{U}_j(t)\rangle}$, where $(i,j)\equiv(1,2,3)$.

\section{Average position of the particle and average work done on the particle}
\label{asymp}
In this section, we show the small- and large-time behaviors of mean position of the particle and mean work on the particle for \fref{fig:ur-x-od}. Notice that these behaviors depend on the choice of the protocol and the observation time.

At small-time  ($\tau\to 0$, where $\tau=\mathcal{T}/t_k$), the mean position of the particle is 
\begin{align}
\overline{\langle x(t) \rangle}=
\begin{cases}
\xrightarrow[]{t=\mathcal{T}} \frac{L \tau}{2}+O(\tau^2)&\text{for}~\lambda(t)=\frac{Lt}{\mathcal{T}},\\
\\
\xrightarrow[]{t=\mathcal{\pi T}} \frac{L \tau}{\pi}[1-\cos(\pi^2)]+O(\tau^2)&\text{for}~\lambda(t)=L \sin \frac{\pi t}{\mathcal{T}},\\
\\
\xrightarrow[]{t=\mathcal{T}} \frac{L \tau}{\pi}+O(\tau^2)&\text{for}~\lambda(t)=\frac{Lt}{\mathcal{T}} \sin \frac{\pi t}{\mathcal{T}},\\
\end{cases}
\label{s-x}
\end{align}
and the mean work done is 
\begin{align}
\overline{\langle \mathcal{W} \rangle}=
\begin{cases}
\xrightarrow[]{t=\mathcal{T}} \frac{L^2}{2 Dt_k}+O(\tau)~ &\text{for}~\lambda(t)=\frac{Lt}{\mathcal{T}},\\
\\
\xrightarrow[]{t=\pi\mathcal{T}} \frac{L^2 [1-\cos(2\pi^2)]}{4 Dt_k}+O(\tau)~&\text{for}~\lambda(t)=L \sin \frac{\pi t}{\mathcal{T}},\\
\\
\xrightarrow[]{t=\mathcal{T}} \frac{L^2\tau [2\pi^2-3]}{12 D \pi^2 t_k}+O(\tau^2)~&\text{for}~\lambda(t)=\frac{Lt}{\mathcal{T}} \sin \frac{\pi t}{\mathcal{T}},
\end{cases}
\label{s-w}
\end{align}
whereas at large-time ($\tau\to\infty$), the mean position of the particle is 
\begin{align}
\overline{\langle x(t) \rangle}=
\begin{cases}
\xrightarrow[]{t=\mathcal{T}} L+O(\frac{1}{\tau})~ &\text{for}~\lambda(t)=\frac{Lt}{\mathcal{T}},\\
\\
\xrightarrow[]{t=\mathcal{\pi T}} L \sin(\pi^2)+O(\frac{1}{\tau})~&\text{for}~\lambda(t)=L \sin \frac{\pi t}{\mathcal{T}},\\
\\
\xrightarrow[]{t=\mathcal{T}} \frac{L \pi}{\tau}+O(\frac{1}{\tau^{2}})\quad \quad &\text{for}~\lambda(t)=\frac{Lt}{\mathcal{T}} \sin \frac{\pi t}{\mathcal{T}},\\
\end{cases}
\label{l-x}
\end{align}
and the mean work on the particle is 
\begin{equation}
\overline{\langle \mathcal{W} \rangle}=
\begin{cases}
\xrightarrow[]{t=\mathcal{T}} \frac{L^2}{Dt_k\tau}+O(\frac{1}{\tau^{2}})~~~~~~~~~~~~~~~~~~ \text{for}~\lambda(t)=\frac{Lt}{\mathcal{T}},\\
\\
\xrightarrow[]{t=\pi\mathcal{T}} \frac{L^2 \pi [2\pi^2+\sin(2\pi^2)]}{4 Dt_k \tau}+O(\frac{1}{\tau^{2}})~\text{for}~\lambda(t)=L \sin \frac{\pi t}{\mathcal{T}},\\
\\
\xrightarrow[]{t=\mathcal{T}} \frac{L^2 (3+2\pi^2]}{12 D t_k \tau}+O(\frac{1}{\tau^{2}})~~~~~~~~~~~\text{for}~\lambda(t)=\frac{Lt}{\mathcal{T}} \sin \frac{\pi t}{\mathcal{T}}.
\end{cases}
\label{l-w}
\end{equation}
From the above equations, it is clear that $\overline{\langle x(t)\rangle}$ approaches to zero and $\lambda(\mathcal{T})$, respectively, at small- ($\tau\to 0$) and large-time $(\tau\to\infty)$. The mean work on the particle approaches to zero for $\lambda(\mathcal{T})=0$, however, it is non-zero for protocols that do not return to its initial value $\lambda(0)=0$ in the limit $\tau\to 0$, whereas it tends to zero in the limit $\tau\to\infty$.


\end{document}